# Ultra-low-field magneto-elastocaloric cooling in a multiferroic composite device

Huilong Hou [1], Peter Finkel [2], Margo Staruch [2], Jun Cui [3,4], Ichiro Takeuchi [1,*]

[1] Department of Materials Science and Engineering, University of Maryland, College Park, Maryland 20742, United States of America

[2] Materials Science and Technology Division, U.S. Naval Research Laboratory, Washington, D.C. 20375, United States of America

[3] Division of Materials Science and Engineering, Ames Laboratory, Ames, Iowa 50011, United States of America

[4] Department of Materials Science and Engineering, Iowa State University, Ames, Iowa 50011, United States of America

* Correspondence and requests for materials should be addressed to I.T. (takeuchi@umd.edu)

**Abstract**

The advent of caloric materials for magnetocaloric, elastocaloric, and electrocaloric cooling is changing the landscape of solid state cooling technologies with potentials for high-efficiency and environmentally-friendly residential and commercial cooling as well as heating applications. Given that caloric materials are ferroic materials which undergo first (or second) order transitions near room temperature, they open up intriguing possibilities for novel multiferroic devices with hitherto unexplored functionalities coupling their thermal properties with different fields (magnetic, electric, and stress) through composite configurations. Here, we demonstrate a composite magneto-elastocaloric effect with ultra-low magnetic field (0.16 T) in a compact geometry to generate a cooling temperature change as large as 4 K using a magnetostriction/superelastic alloy composite. Such composite systems can be used to circumvent shortcomings of existing technologies such as the need for high-stress actuation mechanism for elastocaloric materials and the high magnetic-field requirement of magnetocaloric materials, while enabling new applications such as compact remote cooling devices.

Elastocaloric cooling exploiting the stress-induced martensitic phase transformation of shape memory alloys (SMAs) has recently emerged as a strong alternative-cooling-technology candidate due to the intrinsically high coefficient of performance of elastocaloric materials.[1, 2, 3, 4] Compared to other solid-state cooling techniques, its potentials for high-efficiency cooling systems are only rivaled by magnetocaloric cooling.[5] Previously, compression-based 400 W systems and tension-based device and active regenerators have been demonstrated using the elastocaloric effect.[6, 7, 8, 9] Despite its high efficiency, one disadvantage of elastocaloric cooling is the large stress required to induce the martensitic transformation. For a commonly available Ni-Ti shape memory alloy, for instance, more than 600 MPa is required in compression for the transformation.[10, 11] There are only a handful of engineering options for exerting such large stress, making it challenging to design compact cooling devices. Herein, we enlist magnetostrictive strain to induce elastocaloric cooling in a composite configuration. We demonstrate multiferroic composite devices consisting of $Tb_xDy_{1-x}Fe_2$ (x ~ 0.3, Terfenol-D) which can provide strain with load stress as large as 880 MPa[12] and a Cu–Al–Mn shape memory alloy whose adiabatic temperature change, $\Delta T_{ad}$, can be as large as 12.8 K.[13]

There have been many demonstrations of composite multiferroic effects which take place via elastic coupling between magnetostrictive and piezoelectric materials at their interfaces, and they have been explored for a variety of bulk and thin-film device applications including ultra-high-sensitivity magnetic field sensors,[14, 15, 16, 17] cantilever-based mechanical logic devices,[18, 19] as well as voltage-controlled nanoscale magnetic domain memories.[20] They take advantage of mechanical transduction through strain transfer between materials of similar Young's moduli.[21] In this work, we demonstrate the utility of novel multiferroic cooling devices enabled by elastic coupling of a magnetostrictive material with a superelastic shape memory alloy for the first time.

The functionality of our composite multiferroic devices corresponds to the red arrow path in the modified Heckmann diagram (Figure 1(a)), which includes temperature and entropy as a field and a conjugate response parameter, and it is effectively an alternative route to achieve the magnetocaloric effect (black arrow). A well-known issue of intrinsic magnetocaloric materials is the relatively large magnetic field they require to achieve adiabatic cooling. For instance, magnetic field as large as 2 T is needed to induce $\Delta T_{ad}$ of 5.4 K in gadolinium.[22] In contrast, we use Terfenol-D whose magnetostrictive strain can be as large as 1,800 ppm at less than 1 T to mechanically load Cu–Al–Mn SMA, which undergoes transformation with a relatively small stress of ≈100 MPa. Our compact magneto-elastocaloric (M-eC) devices achieve cooling $\Delta T_{ad}$ of 4 K with 0.16 T, which can open up possibilities for an entirely new class of compact and remote cooling applications.

# Results

Strategy of multiferroic composite to utilize low magnetic field for cooling

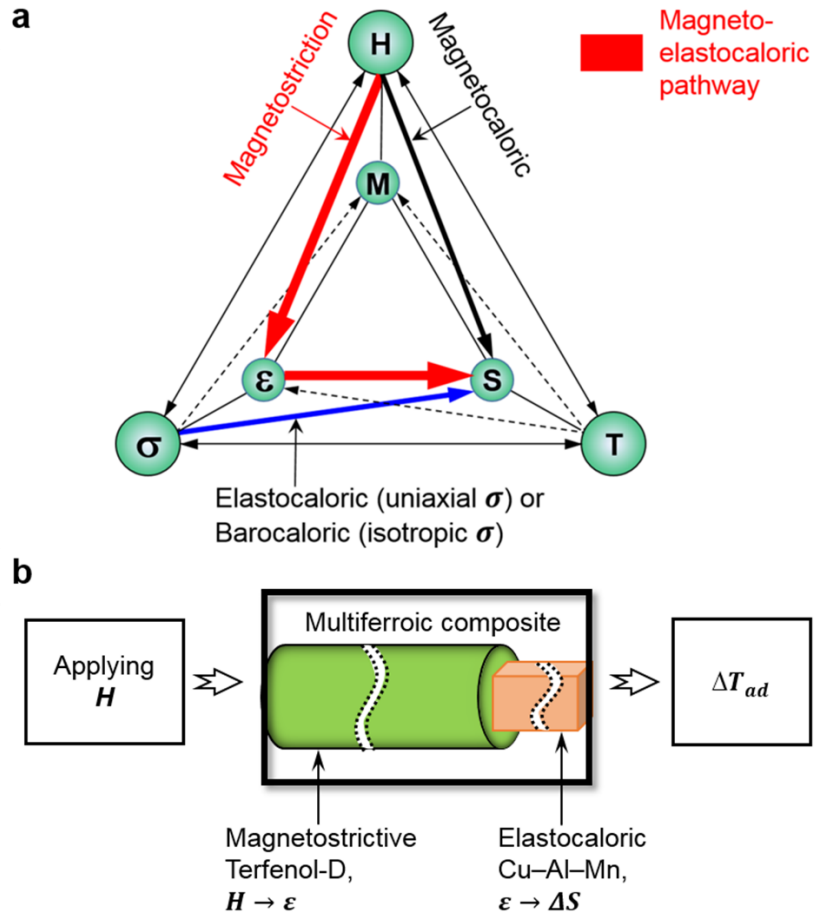

**Figure 1. Magneto-elastocaloric multiferroic composite**. **a**, A modified Heckmann diagram illustrating the pathway leveraging magnetic field for cooling in a multiferroic composite. Symbols: magnetic field ($H$), magnetization ($M$), stress ($\sigma$), strain ($\varepsilon$), temperature ($T$), and entropy ($S$). Red arrow is the composite magneto-elastocaloric pathway demonstrated in this work, while black and blue arrows are intrinsic magnetocaloric and mechanocaloric (elastocaloric and barocaloric) pathways, respectively. **b**, Schematic of the magneto-elastocaloric (M-eC) device, in which a magnetostrictive material (Terfenol-D) and a Cu–Al–Mn shape memory alloy (SMA) are elastically coupled to generate cooling under magnetic field. The Terfenol-D displays extension when magnetic field is applied along the length of Terfenol-D and retraction upon removal of the magnetic field. The SMA generates an isothermal entropy change at a small strain rate and an adiabatic temperature change at a large strain rate through elastocaloric effect.

The schematic of the M-eC device is shown in Figure 1(b). The frame of the device acts as fixed constraints against the overall extension of the multiferroic composite so that mechanical load is transferred from Terfenol-D to Cu–Al–Mn SMA. We use a tabletop electromagnet to generate magnetic field, and a Lakeshore Hall probe is used to measure its magnitude at the surface of Terfenol-D. The temperature change of the Cu–Al–Mn SMA piece in the M-eC device is measured by an infrared camera.

Tuning of cooling induced by low magnetic field

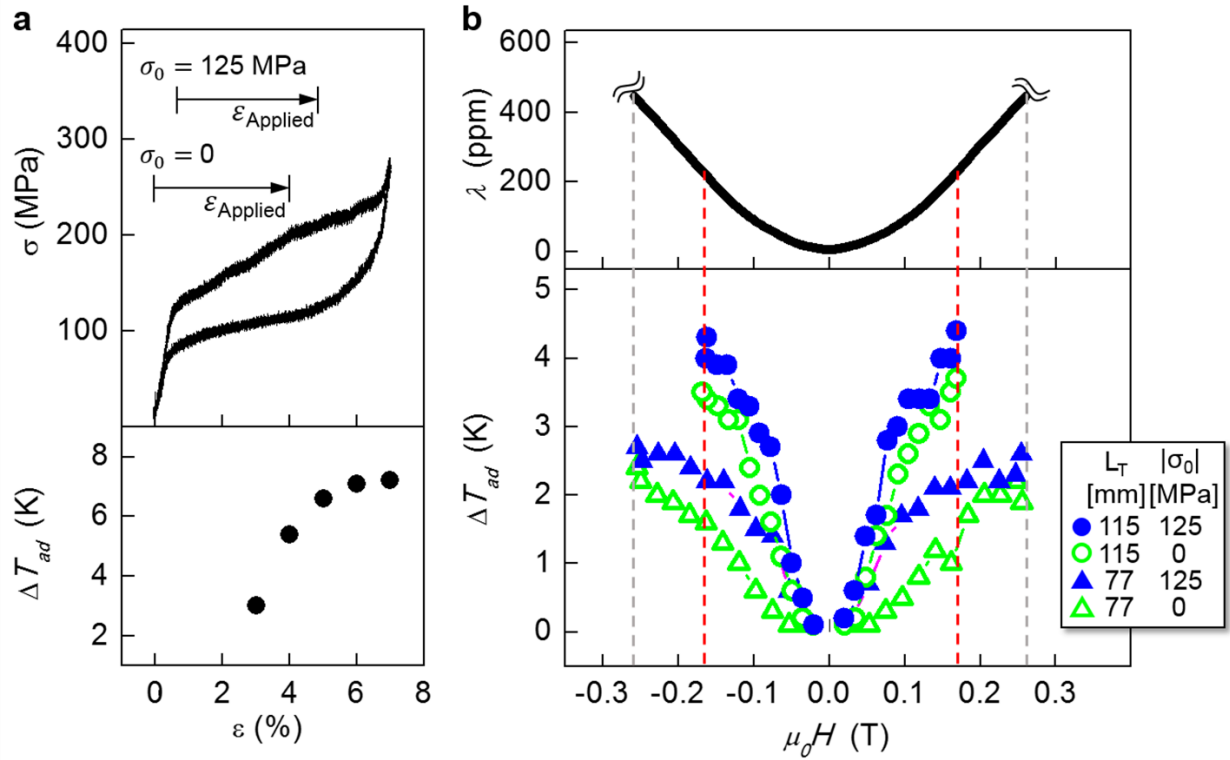

**Figure 2. Tailoring low-field adiabatic cooling. a**, Compressive stress, $\sigma$ at slow loading-unloading (**a**, upper panel) and measured cooling $\Delta T_{ad}$ upon rapid unloading (**a**, lower panel) in the Cu–Al–Mn SMA piece as a function of strain, $\varepsilon$, tested in a servohydraulic load frame. **b**, Magnetostrictive strain, $\lambda$, of the Terfenol-D (**b**, upper panel) and cooling $\Delta T_{ad}$ of the Cu–Al–Mn SMA (**b**, lower panel) in the M-eC device as a function of magnetic field, $\mu_0 H$ (which is removed rapidly for cooling) for two different pre-stress, $\sigma_0$, applied to the SMA for two different lengths, $L_T$, of Terfenol-D used in the device. The pre-stress, $\sigma_0$, signifies the stress level at the onset of applying the strain from magnetostriction of Terfenol-D, $\varepsilon_{Applied}$, to SMA. Magnetostriction of Terfenol-D at various pre-loaded stresses is shown in Supplementary Fig. 1. The dashed lines denote the correspondence of $\Delta T_{ad}$ and $\lambda$ for the same applied field in the M-eC device.

We first investigate the basic properties of components of the M-eC device and then characterize the cooling of the composite device. A typical stress-strain curve of the Cu–Al–Mn SMA piece measured in a conventional servohydraulic load frame is shown in upper panel of Figure 2(a). The strain is fully recoverable, and it is the transformation volume fraction that determines the released (and absorbed) latent heat which in turn controls the ultimate cooling $\Delta T_{ad}$ in the SMA. Different $\Delta T_{ad}$ (lower panel of Figure 2(a)) are attained by rapidly unloading from different strain levels. Terfenol-D displays a maximum magnetostrictive strain of ≈1,800 ppm at 1 T (See Supplementary Fig. 1). Here, we focus on the low-magnetic-field effect: it shows a magnetostriction of ≈450 ppm at ≈0.3 T, which can be used to attain $\Delta T_{ad}$ in the M-eC device as large as 4.4 K (Figure 2(b)). The details of the load transfer in the device is discussed in the Methods section. Out of different Terfenol-D lengths and pre-stress configurations we have looked at, a longer Terfenol-D piece naturally gives rise to a larger strain, and a pre-loaded stress to the Cu–Al–Mn piece also leads to a larger strain, resulting in larger observed $\Delta T_{ad}$.

Attaining adiabatic cooling at low magnetic field

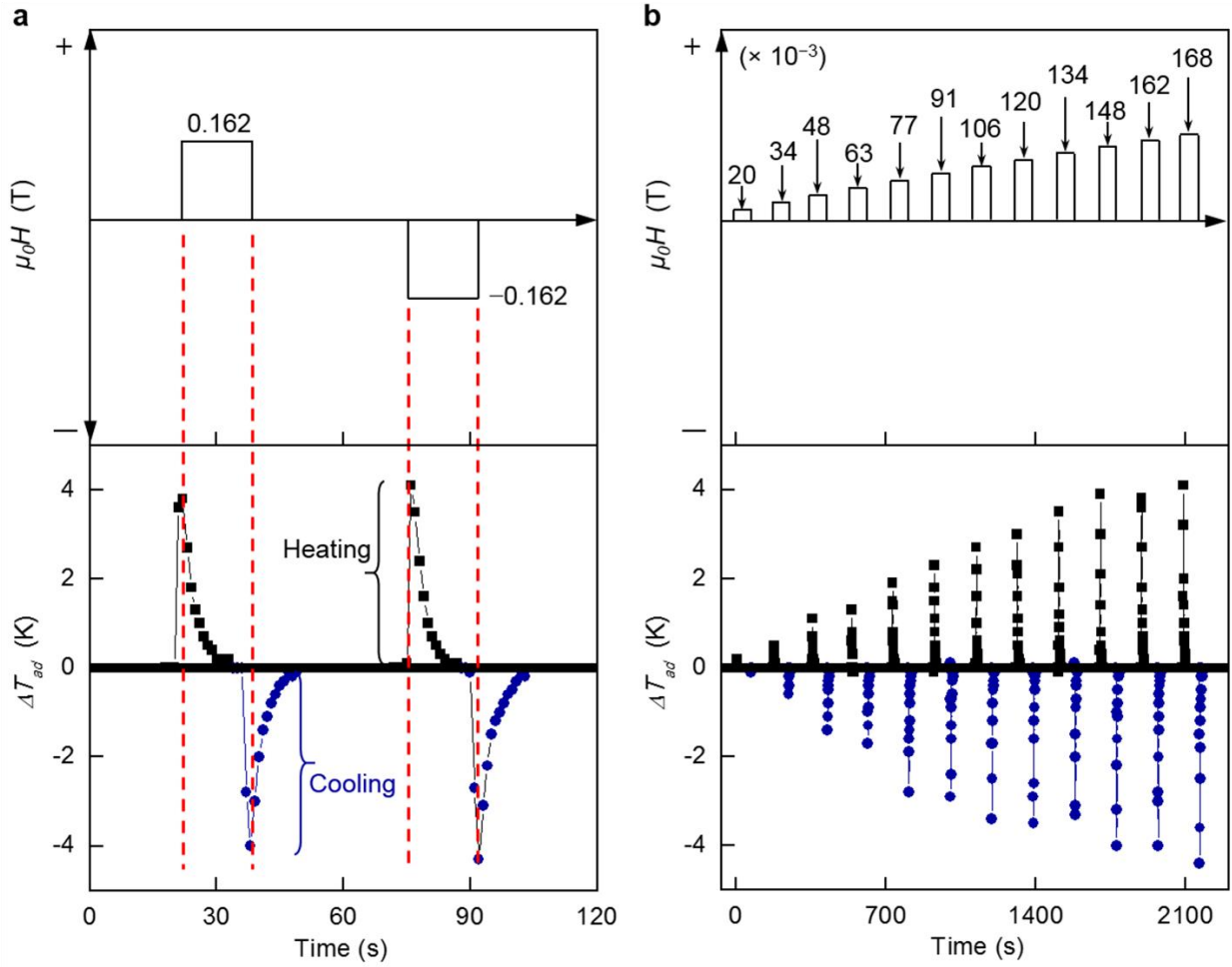

**Figure 3. Adiabatic temperature change observed in the M-eC device under low magnetic fields**. **a**, Waveform of applied magnetic field, $\mu_0 H$, (**a**, upper panel) and corresponding adiabatic temperature change, $\Delta T_{ad}$, of the Cu–Al–Mn SMA in the M-eC device (**a**, lower panel) under rapid application and removal of positive and negative magnetic fields. Black squares indicate the heating part of the data and blue circles mark the cooling part of the data. The red dashed lines in **a** indicate that measured heating $\Delta T_{ad}$ and cooling $\Delta T_{ad}$ are in direct response to the magnetic field change. **b**, Increasing magnitude of magnetic field (**b**, upper panel) and resulting increase in heating and cooling in Cu–Al–Mn SMA (**b**, lower panel) in the M-eC device.

We then preformed a series of measurements of the low-magnetic-field M-eC effect using a 115-mm-long Terfenol-D rod with a pre-loaded stress of 125 MPa (Figure 3). Rapid application and rapid removal of field with positive and negative values (Figure 3(a, upper panel)) leads to a heating and a cooling respectively, both followed by a natural settling back to room temperature (Figure 3(a, lower panel)). The magnetic field of 0.02 T and 0.168 T (Figure 3(b, upper panel)) results in a cooling $\Delta T_{ad}$ (≈heating $\Delta T_{ad}$) of 0.2 K and of 4.1 K respectively (Figure 3(b, lower panel)). $\Delta T_{ad}$ increases linearly with increasing magnitude of magnetic field.

High cooling strength versus low magnetic field

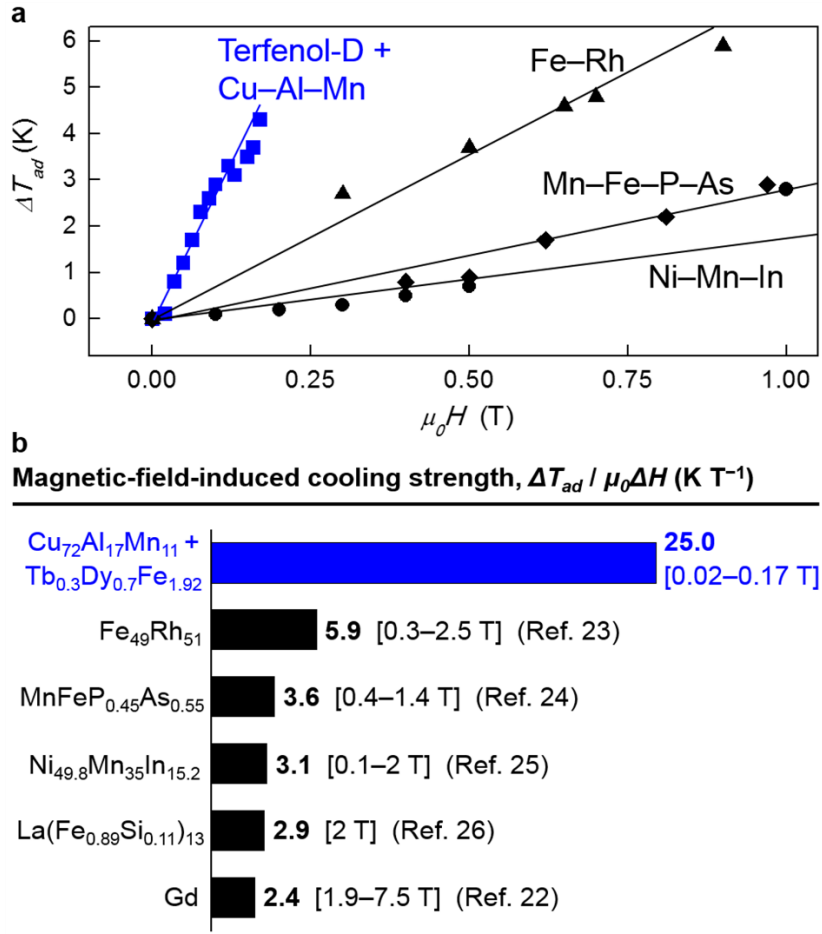

**Figure 4. Comparison of low-field adiabatic temperature change and magnetocaloric cooling strength for selected materials systems. a**, Cooling $\Delta T_{ad}$ as a function of applied magnetic field, $\mu_0 H$. **b**, Comparison of magnetic-field-induced cooling strength defined as the ratio of observed $\Delta T_{ad}$ to the applied magnetic field, $\mu_0 \Delta H$. The value of $\Delta T_{ad}/\mu_0 \Delta H$ for each material is listed on the right side of the bar, and is accompanied by the range of applied magnetic field. Only directly measured experimental data are listed.[22, 23, 24, 25, 26]

Thus, our M-eC devices are functionally able to achieve low magnetic-field-induced cooling. For comparison, we plot $\Delta T_{ad}$ versus applied magnetic field for our devices and selected conventional magnetocaloric materials (Figure 4(a)). Our M-eC devices can attain a $\Delta T_{ad}$ of 4 K with a field of 0.16 T; in contrast, Fe–Rh would require 0.6 T to achieve 4 K, and fields much higher than 1 T are needed for Mn–Fe–P–As and Ni–Mn–In. We define magnetic-field-induced cooling strength to be $\Delta T_{ad}/\mu_0 \Delta H$, and use it as a metric of magnetic-field-induced cooling (Figure 4(b)). Among various materials systems, $\Delta T_{ad}/\mu_0 \Delta H$ of the composite devices here is three times larger than that of Fe–Rh, an intrinsic magnetocaloric material with the highest magnetocaloric magnetic-field-induced cooling strength.[27, 28]

Enabling compactness of cooling devices

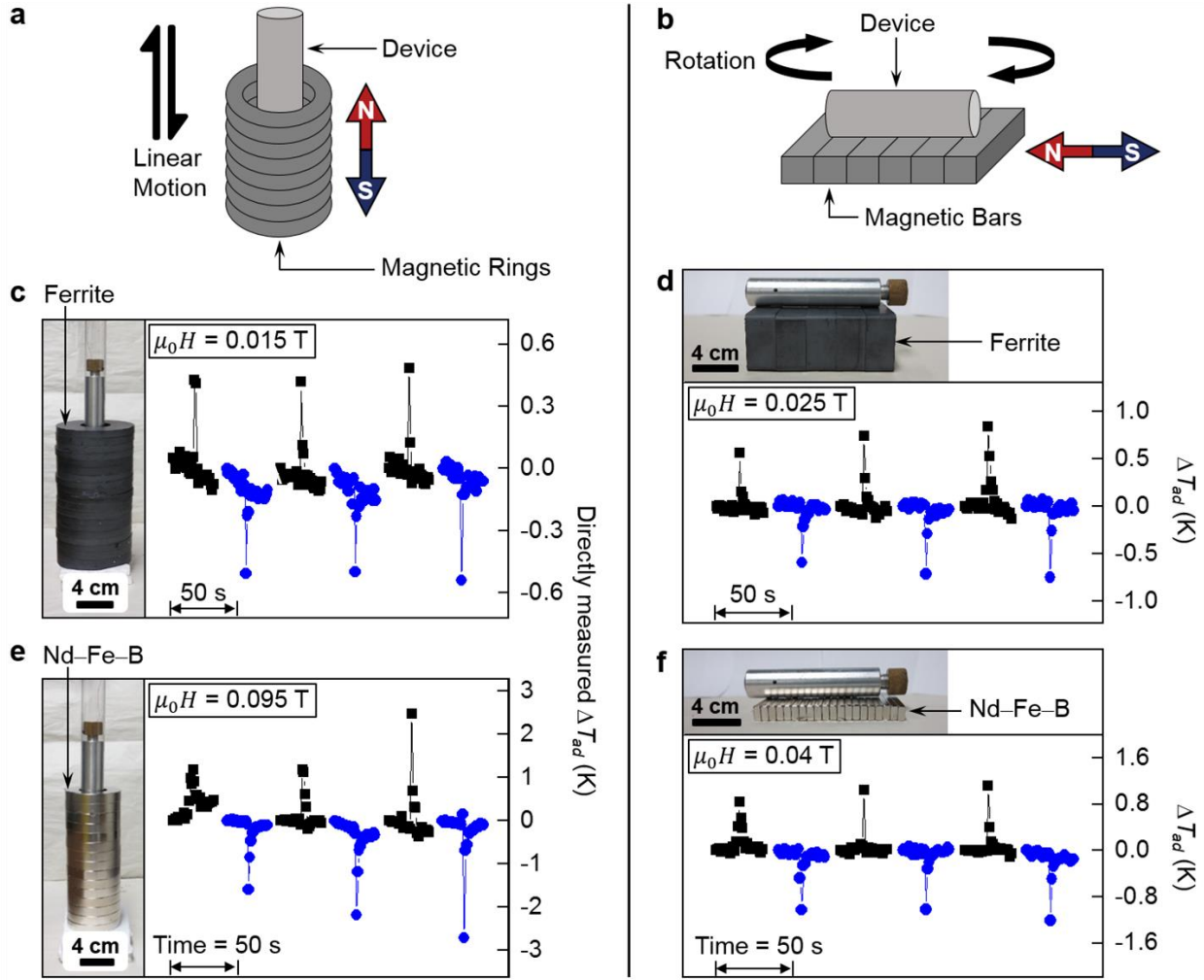

**Figure 5. Demonstrations of compact low-magnetic-field-induced cooling using permeant magnets.** Motion of M-eC device with respect to the magnetic field from stacked ring (**a**) and bar (**b**) permanent magnets in compact geometries leads to heating and cooling. In schematics **a** and **b**, arrows indicate the movement of the device against the static magnet stacks, and the N/S symbols mark the orientation of the magnetic field. Directly measured $\Delta T_{ad}$ is in response to the magnetic field generated by ferrite magnets (**c**, **d**), and by Nd–Fe–B magnets (**e**, **f**). The typical device speed is +/−40 mm s$^{-1}$ for the stacked ring configurations, and 3.5 rad s$^{-1}$ for rotating the device in and out of the field-axis in the bar configuration. In **c**–**f**, the insets are photographs of the M-eC device with permanent magnets, and the fields are the measured values at the surface of Terfenol-D in the M-eC device.

In addition to the high magnetic-field-induced strength, another major advantage of the M-eC devices is their compactness. By enlisting magnetostrictive strain, we have removed cumbersome, large-stress actuation mechanisms to achieve elastocaloric cooling. To illustrate design flexibilities and the straightforward implementation of the M-eC effect, we have demonstrated several geometries where simple relative motion of the device with respect to permanent magnets is used to achieve cooling. In the schematics shown in Figure 5, the device is inserted in and out of stacked ring magnets (Figure 5(a)) or rotated above the plane of magnets in and out of the field-axis (Figure 5(b)). Figure 5(c) and 5(d) show the resulting heating and cooling achieved using ferrite magnets, and Fig. 5(e) and Fig. 5(f) show the magneto-elastocaloric effect achieved with Nd-Fe-B magnets. In all instances, the achieved $\Delta T_{ad}$ is consistent with the magnetic field (insets of Figure 5(c), (d), (e), (f)) measured at the surface of Terfenol-D in the M-eC device.

The demonstrations with the permanent magnets indicate that with proper designs, the M-eC devices can be even more compact. By implementing a design where the permanent magnet is placed even closer to the surface of the Terfenol-D piece, we would be able to attain larger magnetostriction and consequently larger $\Delta T_{ad}$ in the Cu–Al–Mn piece. As an alternative to magnetostriction, it is also possible to use piezoelectric materials to construct another type of multiferroic composites for inducing elastocaloric cooling. In particular, recent development in advanced piezoelectric single crystals[29, 30] indicates that compact piezoelectric/superelastic shape memory alloy composite is a promising option for achieving the "electro-elastocaloric effect."

We envision M-eC devices being deployed for a variety of remote and compact applications including cooling of electronic components, photon detectors, sonar sensors, and micro-refrigerators. An intriguing possibility is to use it for local brain cooling to treat epileptic seizures.[31, 32] We believe the non-contact, wireless nature of the highly-efficient, compact magnetostrictive-SMA composite can be competitive in many application arenas where miniature Peltier cooling devices currently dominate the market. It is important to note that our devices can be operated in the "cooling-only" mode under isothermal loading and adiabatic unloading (by slowly increasing the magnetic field followed by rapid removal as shown in Supplementary Fig. 2) to curtail heating. Naturally, such an operation mode is desirable for a multitude of cooling applications.

## Methods

**Fabrication, characterization, and design of materials.** Copper–Aluminum–Manganese (Cu–Al–Mn) alloys with a nominal composition of $Cu_{72}Al_{17}Mn_{11}$ (at.%) were prepared by induction heating of elemental powders with a purity of 99.9 at.% followed by abnormal grain growth via thermal processing to attain single crystal specimens. Details of the preparations are available in a previous publication.[33] The single crystalline structure was confirmed by x-ray diffraction, and the composition was determined using wavelength dispersive spectroscopy with calibrated standards. The transformation temperatures were analyzed by differential scanning calorimetry, and they were found to be $M_s$ = 270.5 K, $M_f$ = 250.8 K, $A_s$ = 269.5 K, and $A_f$ = 282.8 K. The latent heat was found to be $\Delta H_{A \to M}$ = 4.1 J g$^{-1}$, $\Delta H_{M \to A}$ = 5.3 J g$^{-1}$. Standard stress-strain tests were carried out on a MTS 810 servohydraulic load frame at a strain rate of 0.0002 s$^{-1}$ for isothermal loading-unloading, and at a loading strain rate of 0.0002 s$^{-1}$ and an unloading strain rate of 5 s$^{-1}$ for adiabatic cooling.

Terfenol-D alloy (purchased from ETREMA Products, currently TdVib LLC) had a composition of $Tb_{0.3}Dy_{0.7}Fe_{1.92}$. The linear magnetostrictive strain of the Terfenol-D was estimated 800–1200 ppm. In this work, two Terfenol-D rods with a diameter of 6 mm were machined by electrical discharge machining, one with the length of 77 mm and the other with the length of 38 mm. Both ends of each Terfenol-D rod were fine cut for obtaining smooth surfaces to mechanically interface with Cu–Al–Mn SMA for assembling a multiferroic composite. Magnetostriction under pre-loaded stresses was tested using a MTS 858 load frame with a custom-made sample holder. Micro-Measurements strain gauges mounted on the opposite side of the Terfenol-D were used to determine the strain. (See Reference[34] for details of the apparatus design).

To house the multiferroic composite, a high-strength aluminum frame was customized with an outer diameter of 28.0 mm and an inner diameter of 19.2 mm with two ends capped with Brass knobs, which can be tightened/untightened by an Allen wrench for adding/reducing the pre-stress load to the multiferroic composite. A polyimide ring inside the frame was used to guide the multiferroic composite for avoiding lateral deformation, and ceramic disks were inserted to insulate Cu–Al–Mn SMA from surrounding thermal mass.

The load generated by the Terfenol-D is sufficient to actuate the Cu–Al–Mn SMA. We estimate the 77-mm-long Terfenol-D rod (with a diameter of 6 mm and a cross-sectional area of 28.2 mm$^2$) produces a stress of ≈14 MPa with a magnetic-field-induced extension of 0.092 mm which is constrained by the frame of the M-eC device (Figure 1(b)) resulting in a load of 396 N. This load is able to initiate the phase transformation in the Cu–Al–Mn SMA piece in the M-eC device with a transformation stress of ≈100 MPa, which requires the cross-sectional area of the SMA specimen to be ≈4 mm$^2$ at most. We thus used rectangular Cu–Al–Mn specimens with 2 mm × 1 mm × 2 mm in dimensions: the cross-sectional area which comes in contact with a Terfenol-D rod is 2 mm$^2$, and the 2 mm length is along the crystallographic orientation [110]. In one M-eC device with a 77-mm-long Terfenol-D rod, a strain of 4.6% and a stress of 125 MPa in the prepared Cu–Al–Mn specimens is achieved as a result of the extension of 0.092 mm with the load of 396 N from Terfenol-D. In another measurement, by using a longer Terfenol-D rod (115 mm in length), we are able to achieve a displacement of 0.137 mm resulting in a total compressive strain of 6.9% in the Cu–Al–Mn piece. As the load of 396 N is shared by Terfenol-D and Cu–Al–Mn in the confined frame of the M-eC device, the actual maximum strain that the Cu–Al–Mn specimens experience is estimated to be ≈5%.

**Electromagnet setting and in situ thermal imaging.** Magnetic fields were generated and precisely controlled by an H-frame electromagnet (Micro-Now Instrument Inc.) equipped with a power supply (Model BOP 50-20 MG, Korea Electric Power Corporation) at a maximum output of 50 V and 20 A. The center of the pole caps of the electromagnets was aligned to the longitudinal axis of the composite device, and the space between the pole caps was set to exactly fit the device without gaps to attain highest possible fields to be experienced by the Terfenol-D rods in the device. The magnetic fields were measured using a Lakeshore Hall probe placed in the middle of the device.

During application and removal of magnetic field, the temperature of Cu–Al–Mn piece in the device was directly monitored using an infrared camera (T450sc, FLIR Systems, Inc.) by collecting thermal videos at a frequency of 60 Hz, a spatial resolution of 0.00136 rad, and a thermal sensitivity of 0.03 K at 303 K after calibrations with real-time temperature. A spot meter detected the temperature for an area of $2 \times 2$ mm$^2$ on the Cu–Al–Mn piece in the device during recording of videos, from which thermal images of $320 \times 240$ Pixels were extracted for time-wise analysis of the temperature change. A thin coating of graphite with an emissivity coefficient of 0.95 was sprayed on the surface of Cu–Al–Mn SMA to increase its thermal emissivity. The procedure of recording, extracting, and analyzing was repeated for each combination of the experimental parameters including magnetic field, preloaded stress, and length of Terfenol-D.

**Permanent magnet setting and motion-related cooling.** Commercially available permanent magnets (ring and bar magnets of ferrite and Nd–Fe–B) were used as the source of low magnetic field. The gray ferrite and the silver Nd–Fe–B ring magnets had an outer diameter of 115.3 mm and 76.2 mm, an inner diameter of 44.5 mm and 38.1 mm, and a thickness of 10.2 mm and 12.7 mm, respectively. The gray ferrite and the silver Nd–Fe–B bar magnets had a length of 76.2 mm and 76.2 mm, a width of 50.8 mm and 12.7 mm, and a thickness of 25.4 mm and 6.4 mm, respectively, and they were stacked vertically or horizontally.

**Data availability.** All relevant data are available from the corresponding author upon request.

## Acknowledgements

This work was supported by the Caloric Cooling Consortium, CaloriCool$^{TM}$. The Consortium is a member of the U.S. Department of Energy (DOE) Energy Materials Network, and is supported by the Advanced Manufacturing Office of the Office of Energy Efficiency and Renewable Energy of the U.S. DOE. Advanced Research Projects Agency-Energy (ARPA-E, U.S. DOE) supported the experimentation on shape memory alloys at the University of Maryland under grant number ARPA-E DEAR0000131. The work at Ames Laboratory is supported by the Division of Materials Science and Engineering of the Basic Energy Sciences Programs of the Office of Science of the U.S. DOE under contract number DE-AC02-07CH11358 with Iowa State University. We thank Yoji Yuki and Koutaro Toyotake for providing Cu–Al–Mn raw materials, and Marilyn Wun-Fogel and Nicholas J. Jones for providing a complete set of magnetostriction data of Terfenol-D. We also thank Bao Yang and Zhi Yang for assistance in thermal imaging, and Reinhard Radermacher, Yunho Hwang, and Jan Muehlbauer for discussions on designs of cooling devices.

## Author Contributions

I.T. initiated and supervised the project. P.F., H.H., and M.S. designed and fabricated the device, and characterized the magnetostriction of Terfenol-D alloys. H.H. planned the experiments, prepared Cu–Al–Mn specimens, conducted magnetic-field-induced cooling experiments, and performed data analysis. J.C. contributed to interpretation of data and to comparison of magnetocaloric materials. H.H., I.T., and P.F. wrote the paper with substantial input from other authors. All authors contributed to discussion of the results.

## Additional Information

Supplementary Information is available in the submission.

Competing financial interests: The authors declare no competing financial interests.